**Dirk Helbing**

# New Ways to Promote Sustainability and Social Well-Being in a Complex, Strongly Interdependent World: The FuturICT Approach

*FuturICT is one of six proposals currently being considered for support within the European Commission's Flagship Initiative (see Box 1). The vision of the FuturICT project is to develop new science and new information and communication systems that will promote social self-organization, self-regulation, well-being, sustainability, and resilience. One of the main aims of the approach is to increase individual opportunities for social, economic and political participation, combined with the creation of collective awareness of the impact that human actions have on our world. This requires us to mine large datasets ('Big Data') and to develop new methods and tools: a Planetary Nervous System (PNS) to answer "What is (the state of the world)…" questions, a Living Earth Simulator (LES) to study "What … if…" scenarios, and a Global Participatory Platform (GPP) for social exploration and interaction.*

Today, we understand our physical universe better than our society and economy. Challenges like the financial crisis, the Arab spring revolutions, global flu pandemics, terrorist networks, and cybercrime are all manifestations of our highly and ever more connected world. They also demonstrate the gaps in our present understanding of techno-socio-economic-environmental systems.

In fact, the pace of global and technological change, in particular in the area of Information and Communication Technologies (ICTs), currently outstrips our capacity to handle them. To understand and manage the dynamics of such systems, we need a new kind of science and novel socially interactive ICTs, fostering transparency, trust, sustainability, resilience, respect for individual rights, and opportunities for participation in political and economic processes.

As Columbia University's president Lee C. Bollinger put it: "The forces affecting societies around the world … are powerful and novel. The spread of global systems … are … reshaping our world … raising profound questions."

Information and Communication Technologies (ICTs) are increasing playing a key role for the understanding and solution of problems that our society is facing. Many ICT devices take autonomous decisions, based on real-world data, an internal representation of the outside world, and expectations regarding the future. In some sense, ICT systems are increasingly becoming something like *'Artificial Social Systems'*. Already today, supercomputers perform most financial transactions in the world.

However, today's ICT systems are not constructed in a way that ensures beneficial outcomes. This can result in problems, which we are also facing in our real society, e.g. breakdowns of coordination and performance, 'tragedies of the commons', instabilities, conflicts, (cyber) crime, or (cyber-)war. Furthermore,

ICT systems influence not just their own state, but also impact the real world and human behavior. We, therefore, need a deep understanding of techno-social systems to get ICT systems right and also mitigate our societal problems.

We now have a global exchange of people, goods, money, information, and ideas, which has created a strongly coupled and strongly interdependent world. This often causes feedback and cascading effects, extreme events, and unwanted side effects. In fact, these systems behave fundamentally different from weakly coupled systems. Multi-component systems can be dynamically complex and hardly controllable. Therefore, we need a paradigm shift in our thinking, moving our attention from the properties of the system components to the collective behavior and emergent systemic properties resulting from the interactions of these components.

Note that the paradigm shift from a geocentric to a heliocentric worldview has facilitated many things from modern physics to our ability to launch satellites. In a similar way will the paradigm shift towards an interaction-based, systemic perspective and a co-evolution of ICT with society open up entirely new solutions to address old and new problems, such as financial crises, social and political instabilities, global environmental change, organized crime, the quick spreading of new diseases, and how to build future cities and smart energy systems.

As the previous chapters by Philip Ball have shown, there are promising new approaches to manage complexity: While external control of complex systems is hardly possible due to their self-organized dynamics, one can promote a favorable self-organization by modifying the interaction rules and institutional settings. The potential of a flexible, self-regulating approach has been impressively demonstrated for urban traffic light control and a number of other problems. This approach is based on real-time sensing, short-term anticipation, and the implementation of suitable adaptive interaction rules between the connected system elements. The decentralized self-regulatory principle can be scaled up to systems of almost any size and any kind.

To successfully transfer this approach to other areas and make an effective contribution to mitigating our $21^{st}$ century problems, one needs to develop a better, holistic understanding of the global, strongly coupled and interdependent, dynamically complex systems that humans have created. For this, it is necessary to push complexity science towards practical applicability, to invent a novel data science (that reveals how information is transformed into knowledge and influences human action), to create a new generation of socially interactive, adaptive ICT systems, and to develop entirely new approaches for systemic risk assessment and integrated risk management.

The FuturICT project is a perfect opportunity to foster the creation of such knowledge and the development of the fundaments of new information and communication systems such as a 'Planetary Nervous System' to enable collective, ICT-based awareness of the state of our world, a 'Living Earth Simulator' to explore side effects and opportunities of human decisions, a 'Global

Participatory Platform' to create opportunities for social, economic and political participation, an 'Open Data Platform' (a 'Data Commons') to foster the creativity of people and new business opportunities, a 'Trustable Web' to support safer, privacy-respecting information exchange, as well as value-sensitive ICT to promote responsible interaction. In fact, Europe could well be leading the upcoming age of social and socially inspired innovations, which comes with enormous societal and economic potentials.

**Why FuturICT is needed**

The current lack of a project of this scope and ambition is surprising and, one might even say, deplorable. Big Science projects such as the Human Genome Project, the Large Hadron Collider and the Hubble Space Telescope have revealed, or are revealing, fundamental insights into our genetic constitution and the laws that govern our physical world. However, we have not given the social sciences the same priority as the natural sciences, even though they are highly relevant for maintaining and increasing social well-being. As we have seen in earlier chapters, this is partly because traditional approaches to social problems such as violent conflict and economic instability have often not been very effective in alleviating them. But this is due to the fact that societal challenges have a particular, complex nature as a result of the strongly interconnected patterns and structures of life. Strongly connected, dynamical systems have a number of characteristic properties, for example:

- Even the most powerful computers cannot perform an optimization of the system behaviour in real time, when the number of interacting system elements is large.
- Most real-life complex systems behave probabilistically rather than deterministically, i.e. their behaviour cannot be exactly predicted.
- Strongly connected systems with positive feedbacks tend to change fast, often faster than we can responds and collect enough experience about their behaviour.
- Extreme events occur more often than expected, and can impact the whole system.
- Self-organization and strong correlations dominate the system behaviour. This can lead to surprising, 'emergent' properties of the system.
- The system behaviour can be rich, complex and hard to predict. Planning for the future may be difficult or impossible.
- Complex systems may appear uncontrollable. In particular, opportunities for external, top-down control are very limited.
- Due to possible cascading effects, the vulnerability to random failures or external shocks may be great.
- The loss of predictability and control may lead to an erosion of trust in private and public institutions, which in turn can lead to social, political, or economic destabilization.

Some of these properties challenge our common way of thinking and defy intuition. Even in purely financial terms, the consequences of failing to

appreciate and manage these characteristics of global systems and problems are immense. For example:

- The financial crisis has caused estimated losses of $20 trillion.
- Crime and corruption consumes 2-5% of global GDP: about $2 trillion annually.
- Global military expenditures amount to $1.5 trillion annually.
- The 9/11 terrorist attacks on the US cost the country's economy $90 billion.
- A true influenza pandemic infecting 1% of the world population would cause losses of $1-2 trillion per annum.
- Traffic congestion costs the economy £7-8 billion in the UK alone.

If FuturICT could reduce the impact of these societal problems by just 1%, this would already represent a return of many times the prospective €1 bn Flagship investment. Based on previous success stories regarding a better management of complex systems, an improved understanding of the fundamental underlying issues could actually be expected to facilitate improvements (e.g., in efficiency) of 10-30%.

There are also strong ethical arguments to support the FuturICT project (see Box 2). The fragility of the financial and economic system, for example, carries a serious risk of endangering the stability of society, which may promote crime, corruption, violence, riots, and political extremism, and ultimately undermine democracies and destroy cultural heritage. Rapid scientific progress is needed to learn how to prevent such cascading effects and deterioration. It is also vital to ensure that the social innovations that a project like FuturICT could engender will benefit all of humanity and not end up in the hands of a few stakeholders – a situation that has threatened to arise, for example, in genetic engineering and other transformative technologies.

**Why Information and Communication Technology (ICT) is crucial**

In the global challenges we face, information and communication technologies are part of the problem. People feel that they have created too much speed, too much data, and too much complexity. In fact, our global ICT system is the most complex artefact ever created. It is made up of billions of interacting elements (such as computers, smartphones, users, companies, cars, etc.).

Although humans have built the individual components that compose the system, we are increasingly losing the ability to understand the system as a whole and its interaction with society. No one planned this system as a whole, and no one is in control of it. Often, we do not even know what it 'looks' like – what, for example, the topology of the connectivity network is. We also do not know its weak points and vulnerabilities. We can't predict its behaviour.

But information and communication technologies will inevitably also have to be a part of the solution. We've seen earlier how such systems already collect and embody a tremendous amount of data, some of which encodes vital information about performance, reliability and robustness. Moreover, the pervasiveness of such systems creates an infrastructure that allows us to capture the kind of data, which are needed to model and understand our complex techno-socio-economic-environmental systems. It is an opportunity that must be handled with care (see Box 2).

Despite the need for more data, it's important to collect and use it at the appropriate level. That's to say, FuturICT would not need or desire 'all the data in the world', and the objective is not to model every individual in detail. It is in the very essence of complex systems that this level of detail is not needed: many of the important behaviours, such as trends, norms and cultural shifts, are collective ones, which can be understood without knowing everything about the single individuals and their interactions. That is precisely what makes the complexity approach tractable: it does not involve a 1:1 mapping of the world onto models. In general, different questions require one to focus on different levels of hierarchically organized systems or strongly connected parts of a system, with a degree of data aggregation that is fit to the purpose. (Doctors do not need genetic information to fix a tooth, nor brain scans to operate on the knee.)

Concomitant with the need for massive data collection, FuturICT will require innovations in the extraction of information and meaning. Some new technologies now supply means of gathering data in volumes and rates that exceed our ability to store, retrieve, catalogue, and interpret them. This is a problem felt particularly keenly in the science of genomics and bio-informatics, where data collection has sometimes proceeded apace in the absence of a conceptual framework for asking questions and testing hypotheses. Data is good only to the extent that it can be mined for meaning. To generate valuable knowledge, data mining must often be combined with theoretical models. Advances in this area will, therefore, require input from computer scientists and specialists in data visualization. But the perspective of the social and complexity sciences to identify meaningful trends and correlations will also be essential.

**The components of FuturICT**

The FuturICT project primarily aims at bringing data, models and people together. It will develop new information and communication technologies (ICTs) to collect massive data sets and mine them for useful or meaningful information, and build ICT systems that have the capacity to self-organize and adapt to the collective needs of users. These ICT systems will be the basis of the *FuturICT Platform*, which will have three main components: the '*Planetary Nervous System'*, the '*Living Earth Simulator'*, and the '*Global Participatory Platform'*. The measurement, modelling, and participatory elements of these components will be used for practical applications, such as *'Exploratories' for Society, Economy, Technology*, and the *Environment*. These Exploratories serve to interactively explore interdependencies in our world and will be created by connecting *'Interactive Observatories'* for social well-being, for conflicts and wars, for financial systems, transportation and logistic systems, any many other areas.

The *Living Earth Simulator* will enable the exploration of possible future scenarios at different degrees of detail, employing a variety of perspectives and methods (such as sophisticated agent-based simulations and multi-level models). It will act as a 'Policy Simulator' or 'Policy Wind Tunnel', enabling one to test alternative choices and different policies in advance to explore their possible or likely consequences. These simulations will be enabled by the so-called *World of Modelling* – an open-software platform, comparable to an App store, to which scientists and developers can upload theoretically informed and empirically validated modelling components that map parts of our real world. Rather than giving an ultimate answer, the pluralistic approach of this platform will offer multiple perspectives on difficult problems and, thereby, support better informed decision-making guided by the values and priorities of the respective users.

The Living Earth Simulator will require the development of interactive, decentralized, scalable computing infrastructures, coupled with the access to Big Data. Gathering these data is the role of the *Planetary Nervous System (PNS)*. Our bodies are constantly responding and adjusting to new data – for example, tuning our metabolism to the energy requirements, ambient temperature and current energy reserves. Sensory feedback is essential for navigating our environment, avoiding danger and performing fine motor tasks. The Planetary Nervous System will provide the same sort of function for our Earth. It will be comprised of a global sensor network, where 'sensors' include anything able to provide data in real-time about socio-economic, environmental or technological systems, including the Internet. Such an infrastructure will enable real-time data mining ('reality mining'), e.g. of news, information feeds, or search trends.

A crucial part of the FuturICT platform are the *Interactive Observatories*, which will use the Planetary Nervous System to spot potential weaknesses or problems arising in specific sectors, such as the financial system or conflict-prone regions. These Observatories will continuously monitor the 'health' of the economy, of urban and transportation systems, or of international relations, say, to give early warnings of hazard, where possible. Analogous to seismic-monitoring networks, they will watch out for the build-up of stresses or for precursory signals of impending collapse or catastrophe. They will thus facilitate pre-emptive searches for solutions based on specific, locally tuned scenarios.

The *Global Participatory Platform (GPP)* will make FuturICT's new methods and tools available for everyone (with reputation and transparency mechanisms in place to foster responsible use). This will enable people to look at all interesting issues from many angles and to use the power of crowd sourcing and the wisdom of crowds. The Global Participatory Platform will promote communication, coordination, cooperation and the social, economic and political participation of citizens. In this way, the traditional separation between users and providers, or customers and producers will be overcome, thereby unleashing new economic potentials. Building on the successful principles of Wikipedia and the Web x.0, societies will be able to harness the knowledge and creativity of many minds much more effectively than we can do today.

The Global Participatory Platform will include *Interactive Virtual Worlds*, where possible futures can be explored and enacted using techniques like those developed for multi-player online games. The purpose of these virtual copies of our world is to explore possible futures, i.e. to identify likely systemic outcomes of interactions, given certain 'rules of the game' and institutional settings. For example, one could study different kinds of financial architectures and the market dynamics resulting from them. Such participatory experiments could also inform the designs of shopping malls, airports, and future cities.

In addition to the interconnected systems forming the Living Earth Platform, FuturICT will create an *Innovation Accelerator (IA)* to identify inventions and innovations early on, to distil valuable knowledge from a flood of information, to find the best experts for projects, and to fuel distributed knowledge generation by 'crowd-sourcing'. This Innovation Accelerator will also catalyse the integration of project activities in one single platform. Because the Innovation Accelerator will support communication and flexible coordination in large-scale projects, it will form the basis of the innovative management of the FuturICT Flagship itself.

**Towards more resilient and sustainable systems: How it all comes together**

One of the scientific challenges behind attempts to promote social well-being will be to measure the relevant factors for it on a global scale and in real-time with sufficient accuracy.[1]

Measuring social well-being is even more difficult than measuring GDP. Currently, reliable official numbers for GDP are published only with a delay of many months. However, new ways of measuring GDP have recently been suggested. For example, it seems feasible to estimate GDP from satellite pictures of global light emissions. Such estimates are possible almost in real-time. Similarly, it has been shown that health-related indicators (such as the number of patients during flu pandemics) can be well estimated based on Google trends data.

Therefore, the vision of FuturICT is to make the different dimensions of social well-being globally measurable in real-time. This could be done by mining freely available data on the internet, by sentiment analysis of tweets and blogs, or by use of sensor data of various kinds. Recent attempts to measure happiness and its variation in space, time and across social communities point the way for this.

**Determining the 'social footprint' to protect the fabric of society**

FuturICT's Planetary Nervous System will provide the methods and tools to measure human activities and socially relevant variables in real-time, on a global scale, and in a privacy-respecting way. By extending measurements to social and economic domains, FuturICT will complement and go beyond the scope of similar projects focused on environmental and climate-oriented measurements (e.g. 'Planetary Skin' and 'Digital Earth'). The ambition of the Planetary Nervous System is more than measurement. It also intends to create individual and collective awareness of the impact of human decisions and actions, particularly on the social fabric on which our society is built (the 'social footprint').

---

[1] Note that delayed policy response may cause an unstable system dynamics.

The Planetary Nervous System serves to detect possible opportunities and threats, in order avoid mistakes that one may regret later on. This requires a certain ability to anticipate (in a probabilistic way) possible courses of events. While our own consciousness performs such anticipation by 'mental simulation', FuturICT's Living Earth Simulator will perform the equivalent task for complex techno-socio-economic-environmental systems, by simulating simplified models of our society and economy and other relevant activities in our world. On top of this, Interactive Virtual Worlds and Mixed Reality Environments will provide an online laboratory to explore human interactions under close-to-realistic conditions.

While FuturICT's new concepts for the measurement, simulation and interactive exploration of the impact of human decisions and actions will support collective awareness, the Global Participatory Platform will create new opportunities for social, economic, and political participation. Altogether (and with suitable rating and reputation mechanisms in place), this will promote better decisions and responsible actions. In particular, measuring the value of human and social capital and quantifying the 'social footprint' will help us to protect the social fabric on which our society is built, in a similar way as the measurement of the 'environmental footprint' has empowered people and institutions to better protect our environment.

**In conclusion**

The FuturICT flagship project seeks to create an open and pluralistic, global but decentralized, democratically controlled information platform that will use online data together with novel theoretical models to achieve a paradigm shift in our understanding of today's strongly interdependent and complex world and make both our society and global ICT systems more flexible, adaptive, resilient, sustainable, and humane through a participatory approach. FuturICT is a big project: an unprecedented multi-disciplinary, international scientific endeavour requiring the collaborative effort of hundreds of scientists worldwide (see Box 3). The first practical results are expected 2-3 years after the project starts, with results and tools being made available to the public throughout the ten-year lifetime of the project.

This is an ambitious goal, but one that is within our reach. It will require advances in ICT, social science and complexity science. It will involve all levels and facets of society in developing and shaping the project's outcomes. Without an enterprise of this sort, the world is sure to fall increasingly behind the reach of orderly planning and management. Given the 21$^{st}$ century challenges that we need to tackle, it would be irresponsible *not* to undertake a project of this ambition. Everyone in the world is being affected by these new problems. They should all have a say in their solution.

**Further Reading**

FURTHER INFORMATION

**Box 1: The FET Flagship Initiative**

The FuturICT project is the response to the Flagship Initiative launched by the Future and Emerging Technology (FET) section of the European Commission: a call for 'Big Science' projects with genuinely transformative potential and a 'man on the moon' scope of vision. In the first round of the process, 21 candidates were narrowed down to six Flagship Pilots, of which FuturICT received the highest rating.

Each pilot will submit detailed proposals in April 2012, and at least two of them will be selected for funding of up to €1 billion each over ten years, starting by the end of 2013. (Note that this is about a tenth or less of what is currently invested in other Big Science projects such as the Large Hadron Collider at CERN, the ITER nuclear-fusion reactor, the Galileo satellite program and the Human Genome Project.) Approximately half of the money must be mobilized by the project partners from national budgets and funding agencies, from business and industry, or from donations. A considerable fraction of the Flagship funding will be distributed through 'Open Calls', which will allow a wide scientific community to contribute to the goals.

Among the particular strengths of the FuturICT proposal are:
- its societal relevance,
- the immediate importance of its results for the everyday life of ordinary citizens,
- its large and rapidly growing community and multidisciplinary nature,
- the participation of many European countries,
- the significant support of scientific communities in other continents,
- its open-access and transparent project architecture, and
- its educational activities.

**Box 2: The FuturICT Approach**

FuturICT has also a strong ethical motivation. Among its aims are

- to promote human well-being and responsible behaviour,
- to promote the provision of unbiased, high-quality information, and to increase individual and collective awareness of the impact of human behaviour,
- to reduce vulnerability and risk, increase resilience, and reduce damages,
- to develop contingency plans and explore options for future opportunities and challenges,
- to increase sustainability,
- to facilitate flexible adaptation,
- to promote fairness and happiness,

- to protect and increase social capital,
- to improve opportunities for economic, political, and social participation,
- to find a good balance between central and decentralised (global and local) control,
- to protect privacy and other human rights, pluralism and socio-bio-diversity,
- to support collaborative forms of competition ('co-opetition') and
- to promote responsible behaviour.

Over its ten-year funding period, FuturICT will aim to develop a new ICT paradigm, focusing on socio-inspired ICT, the design of a 'trustable web', ethical, value-sensitive, culturally fitting ICT (responsive+responsible), privacy-respecting data-mining technologies that give users control over their own data, platforms for collective awareness, a new information ecosystem, the co-evolution of ICT with society, public involvement and democratic control. FuturICT plans to provide an open data, simulation, exploration and participatory platform to promote new opportunities for everyone. This platform would represent a new public good on which all kinds of services can be built. It will support both commercial and non-profit activities. To prevent misuse and enable reliable high-quality services, the platform will be decentralized and built on principles of transparency, reputation and self-regulation.

**Box 3: FuturICT's partners and supporters**

FuturICT is currently supported by about 1000 scientists worldwide (see www.futurict.eu). It involves Europe's academic powerhouses, such as ETH Zurich, University College London (UCL), Oxford University, the Fraunhofer Society, the Consiglio Nazionale delle Ricerche (CNR), the Centre National de la Recherché Scientifique (CNRS), Imperial College, and many more excellent academic institutions. More than five supercomputing centres support FuturICT. The project has letters of support from the OECD, the European Commision's Joint Research Center (JRC), research labs of Disney, IBM, Microsoft, SAP, Xerox, and Yahoo, international companies including banks, insurance and telecommunication companies, several regulatory authorities, and notable individuals such as George Soros. Furthermore, FuturICT has already managed to integrate many different research communities, and its leaders have a long track record of successful collaborations between scientists across disciplinary boundaries. FuturICT's supporters have also been involved in hundreds of successful projects with business partners.